\documentclass[12pt,epsf]{article}
\usepackage{graphicx}
\begin{document}
%%%%%%%%%%%%%%%%%%%%%%%%%%%%%%%%%%%%%%%%%%%

\def\a{\alpha}
\def\b{\beta}
\def\c{\varepsilon}
\def\d{\delta}
\def\e{\epsilon}
\def\f{\phi}
\def\g{\gamma}
\def\h{\theta}
\def\k{\kappa}
\def\l{\lambda}
\def\m{\mu}
\def\n{\nu}
\def\p{\psi}
\def\q{\partial}
\def\r{\rho}
\def\s{\sigma}
\def\t{\tau}
\def\u{\upsilon}
\def\v{\varphi}
\def\w{\omega}
\def\x{\xi}
\def\y{\eta}
\def\z{\zeta}
\def\D{\Delta}
\def\G{\Gamma}
\def\H{\Theta}
\def\L{\Lambda}
\def\F{\Phi}
\def\P{\Psi}
\def\S{\Sigma}

\def\o{\over}
\def\beq{\begin{eqnarray}}
\def\eeq{\end{eqnarray}}
\newcommand{\gsim}{ \mathop{}_{\textstyle \sim}^{\textstyle >} }
\newcommand{\lsim}{ \mathop{}_{\textstyle \sim}^{\textstyle <} }
\newcommand{\vev}[1]{ \left\langle {#1} \right\rangle }
\newcommand{\bra}[1]{ \langle {#1} | }
\newcommand{\ket}[1]{ | {#1} \rangle }
\newcommand{\EV}{ {\rm eV} }
\newcommand{\KEV}{ {\rm keV} }
\newcommand{\MEV}{ {\rm MeV} }
\newcommand{\GEV}{ {\rm GeV} }
\newcommand{\TEV}{ {\rm TeV} }
\def\diag{\mathop{\rm diag}\nolimits}
\def\Spin{\mathop{\rm Spin}}
\def\SO{\mathop{\rm SO}}
\def\O{\mathop{\rm O}}
\def\SU{\mathop{\rm SU}}
\def\U{\mathop{\rm U}}
\def\Sp{\mathop{\rm Sp}}
\def\SL{\mathop{\rm SL}}
\def\tr{\mathop{\rm tr}}

\def\IJMP{Int.~J.~Mod.~Phys. }
\def\MPL{Mod.~Phys.~Lett. }
\def\NP{Nucl.~Phys. }
\def\PL{Phys.~Lett. }
\def\PR{Phys.~Rev. }
\def\PRL{Phys.~Rev.~Lett. }
\def\PTP{Prog.~Theor.~Phys. }
\def\ZP{Z.~Phys. }

%%%%%%%%%%%%%%%%%%%%%%%%%%%%%%%%%%%%%%%%%%%%%%%%%%%%%%%%%%%%%%%%%%%%

\baselineskip 0.7cm

\begin{titlepage}

\begin{flushright}
UT-06-20
\end{flushright}

\vskip 1.35cm
\begin{center}
{\large \bf

 A New Inflation Model with Anomaly-mediated Supersymmetry Breaking

}
\vskip 1.2cm
M. Ibe${}^{1}$, Y.  Shinbara${}^{1}$ and T. T. Yanagida${}^{1,2}$
\vskip 0.4cm

${}^1${\it Department of Physics, University of Tokyo,\\
     Tokyo 113-0033, Japan}

${}^2${\it Research Center for the Early Universe, University of Tokyo,\\
     Tokyo 113-0033, Japan}

\vskip 1.5cm

\abstract{
If there are a large number of vacua, multi-inflation may be a more 
mediocre phenomenon rather than a single inflation.
In the multi-inflation scenario, new inflation is most likely the last inflation, since its
energy scale is naturally low.
Furthermore, it may explain the observed spectral index of the cosmic microwave 
background radiations.
We show, in this letter, that a new inflation model proposed in supergravity accounts for all the 
present observations
assuming anomaly mediation of supersymmetry breaking. 
As a result, we find that the relic density of the winos is consistent with the observed dark 
matter density
in a wide range of the wino mass, $100$\,GeV$\lsim m_{\tilde{w}} \lsim2$\,TeV, albeit for a low
reheating temperature $T_{R}\simeq 10^{6-7}$\,GeV.
 }
\end{center}
\end{titlepage}

\setcounter{page}{2}

\section{Introduction}

The existence of a large number of vacua is the most exciting discovery
of string theory \cite{Bousso:2004fc}. This multiplicity of vacua called as 
"landscape of vacua" \cite{Susskind:2003kw} provides a theoretical basis for 
the anthropic explanation on the small cosmological constant \cite{Weinberg:1987dv}.
That is, if the cosmological constant, $\Lambda_{\rm cos}$, takes a wide 
range of values in the full vacua, one may find intelligent observers in 
sub-vacua  where the $\Lambda_{\rm cos}$ takes a sufficiently small value 
for the presence of the observers. In this landscape the flat potential of 
a scaler field for inflation is also naturally explained, since it is necessary for
the observers to exist. Furthermore, multi-inflation seems a more mediocre phenomenon
in the landscape, rather than a single inflation \cite{Vilenkin:1994ua,Ibe:2004mp}.

If the multi-inflation takes place, a lower-scale inflation starts at a later
time. Thus, the last inflation we see today is most likely a low-scale inflation.
We consider a new inflation as the last one, since new inflation is known to have naturally 
a low-scale Hubble constant.
In a recent article~\cite{Ibe:2006fs} we have shown that a new inflation 
model~\cite{Kumekawa:1994gx,Izawa:1996dv} 
constructed in the supergravity (SUGRA) predicts the spectral index $n_s$ of the cosmic
microwave background radiations as $n_s\simeq 0.95$~\cite{Ibe:2006fs,Izawa:2003mc} in a large parameter
space. This result turns out to be
well consistent with the recent WMAP observation~\cite{Spergel:2006hy}, 
$n_{s} = 0.951^{+0.015}_{-0.019}$ (68\%C.L.). 
Encouraged by this success, we examine, in the present letter, if this new inflation model is 
consistent with all other observations. We assume anomaly-mediation models for supersymmetry
(SUSY) breaking, since gravity-mediation models suffer from a serious 
gravitino-overproduction problem \cite{Kawasaki:2006gs, Kawasaki:2006hm,Ibe:2006am,ISY2}. 
We stress, in particular, that the new inflation model (with anomaly-mediated SUSY breaking)
may explain the observed dark matter density as well as the  baryon asymmetry in the universe.
As we show, the relic density of the wino LSP is consistent with the observed dark matter density 
in a wide range of the wino mass, $100$\,GeV$\lsim m_{\tilde{w}} \lsim2$\,TeV, albeit for a low 
reheating temperature $T_{R}\simeq 10^{6-7}$\,GeV.

We also briefly discuss gauge-mediation models in the last section. 

\section{A new inflation model}
\label{sec:newinflation}
Let us discuss a new inflation model 
considered in Ref.~\cite{Kumekawa:1994gx,Izawa:1996dv}. 
In the model, the superpotential and the K\"ahler potential of an inflaton chiral superfield $\phi$ are given by
\begin{eqnarray}
W_{\rm inf} = v^{2} \phi - \frac{g}{n+1} \frac{\phi^{n+1}}{M_G^{n-2}},
\label{eq:Super}
 \end{eqnarray}
and 
\begin{eqnarray}
K_{\rm inf} = |\phi|^{2} + \frac{k}{4} \frac{|\phi|^{4}}{M_G^2} + \cdots,
\label{eq:Kahler}
\end{eqnarray}
respectively.
Here,  $v^{2}$ denotes a dimensionful parameter and 
$g$ and $k$ denote dimensionless coupling constants.
We take the parameters $v^{2}$ and $g$ positive without a loss of  generality. 
And $n$ is an integer number greater than 2.
Hereafter, we take the unit with the reduced Planck scale, $M_{G}\simeq 2.4\times 10^{18}$ GeV,
equal to one.
The above superpotential is generic under a discrete $Z_{2 n}$  $R$-symmetry 
with $\phi$'s charge 2.
Then, the effective scalar potential of the inflaton $\varphi = \sqrt{2} {\rm Re}[\phi]$ is well
approximated by
\begin{equation}
 V(\varphi) \simeq v^4 - \frac{k}{2}v^4 \varphi^2
 -\frac{g}{2^{\frac{n}{2}-1}}v^2\varphi^n
 +\frac{g^2}{2^n}\varphi^{2n},
\label{eq:potential}
\end{equation}
for the inflationary period near the origin, $\varphi=0$.
This potential is very flat for $n\geq3$ and $|k|\ll 1$.
In the following, we consider that the previous inflation drives the inflaton $\varphi$ to
the origin~\cite{Izawa:1997df}, and 
assume $k>0$ so that the inflaton $\varphi$ rolls down
slowly to the potential minimum from near the origin.
Note that the inflaton obtains a mass,
\begin{equation}
\label{eq:phimass}
m_{\varphi} \simeq n g \phi_{0}^{n-1} \simeq n v^{2} \left(\frac{v^2}{g}\right)^{-\frac{1}{n}},
\end{equation}
at the potential minimum,
\begin{equation}
 \phi_0 = \frac{1}{\sqrt{2}}\varphi_{0}\simeq \left(\frac{v^2}{g}\right)^{\frac{1}{n}}.
 \label{eq:phivev}
\end{equation}

We should stress here that the present inflation model contains only four parameters, $v,k,g$ and $n$.
We now show all parameters are determined by the observations, provided that the 
SUSY is broken at low energies, 
that is, the gravitino mass $m_{3/2} < 10^6$ GeV. 

First of all, we should note one of the most remarkable features of the present new inflation model;
the inflation scale $v$ is directly related to the gravitino mass~\cite{Kumekawa:1994gx}.
The important point is that a constant term in the superpotential is generated in the true minimum
of the inflaton potential. Thus, the negative energy at the inflaton potential minimum 
is to be canceled out by the positive energy $\Lambda^4_{\rm{SUSY}}$ induced by  the SUSY breaking to have the cosmological constant vanishing. Thus,
 we have a condition as
\begin{equation}
 \Lambda^4_{\rm{SUSY}}-3|W_{\rm inf}(\phi_0)|^2=0.
\end{equation}
Then, the gravitino mass is given by
\begin{equation}
 m_{3/2} = W_{\rm inf}(\phi_0) 
 \simeq
 \frac{n v^2}{n+1}
  \left(\frac{v^2}{g}\right)^{\frac{1}{n}}.
  \label{eq:gmass}
\end{equation}
We see that the gravitino mass is basically given by the inflation scale $v$.

\begin{figure}[t!]
 \begin{center}
 \includegraphics[width=6.5cm]{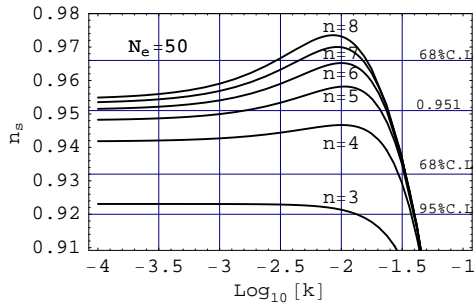}
 \caption{The $k$ dependence of the spectral index $n_s$ for $n=3-8$ and $N_{e}=50$.  
 The horizontal grid lines correspond to the result of WMAP three year data~\cite{Spergel:2006hy}. 
 For $k\gsim10^{-2}$, $n_{s}\simeq 1-2k$, 
 and for $k=0$, $n_s \simeq (N_{e}(n-2)-(n-1))/(N_e(n-2)+(n-1))$.}
 \label{fig:spectrum1}
 \end{center}
\end{figure}

Before determining the inflation scale $v$, we derive constraints on the parameters $n$ and $k$.
The spectral index of the density fluctuations is given 
by~\cite{Izawa:2003mc,Ibe:2006fs,Izawa:1996dv}
\begin{eqnarray}
 n_s &\simeq& 1-6\epsilon + 2 \eta,\\ 
  &\simeq& 1 - 2 k \left[1 + \frac{n-1}
 {\left\{1+\frac{k}{1-k}(n-1)\right\}e^{{N_e}k(n-2)}-1}\right],
 \label{eq:spectrum}
\end{eqnarray}
where $\epsilon$ and $\eta$ denote  so-called slow roll parameters at the horizon crossing,
and $N_{e}$ is the e-folding number of the present universe~\cite{Lyth}.
Note that the spectral index depends on neither $v^2$ nor $g$ explicitly and is mainly 
given by the parameter $k$.
As stressed in the introduction, the present new inflation model predicts the spectral 
index $n_s\simeq 0.95$ for $n\geq 4$ and $k\lsim 10^{-2}$ 
(see Fig.~\ref{fig:spectrum1}),
which is well consistent with the recent WMAP result,
$n_{s} = 0.951^{+0.015}_{-0.019}$ (68\%C.L.)~\cite{Spergel:2006hy}.%
\footnote{As claimed in Refs.~\cite{Lewis:2006ma,Seljak:2006bg,Kinney:2006qm},
the constraint on the spectral index is somewhat relaxed, and especially, 
the spectrum with $n_{s}=1$ is marginally inside the 95\% C.L. region.} 
The WMAP observation favors models with $n\geq4$ and $k\lsim 10^{-2}$.
We, therefore, take $n\geq4$ and $k\lsim 10^{-2}$ in the the following discussion.

%%%%%%%%%%%%%%%%%%%%%%%%%%%%%%%%%%%%%%%%%%%
%%%%%%%%%%%%%%%%%%%%%%%%%%%%%%%%%%%%%%%%%%%

We now come to the point to determine the inflation scale $v$. 
The inflation scale $v$ is given by 
\begin{eqnarray}
\label{eq:appv}
v&\simeq& \left( 5\sqrt{6}\pi n \d  \right)^{\frac{n-2}{2n-6}} 
\left( \frac{1}{n(n-2)N_{e}}\right)^{\frac{n-1}{2n-6}}
\left( \frac{1}{g} \right)^{\frac{1}{2n-6}},
\end{eqnarray}
for $n\geq 4$ and $k\lsim 10^{-2}$, where
we have neglected a weak dependence on $k$ (see Eq.~(19) in Ref.~\cite{Ibe:2006fs} for details).%
\footnote{For the model with $n=3$, the density fluctuations do not determine the inflation scale $v$,  but it determines the coupling constant $g$ as $g \sim 10^{-(6-7)}$.}
Here, the parameter $\d$ is  the amplitude of the density fluctuations which is
measured~\cite{Spergel:2006hy} as
\begin{eqnarray}
\label{eq:COBE}
\d = \frac{1}{5 \sqrt{3}\pi}\frac{V^{3 \over 2}}{|V^\prime|} \simeq 1.9 \times 10^{-5}.
\end{eqnarray}
%%%%%%%%% NOTE %%%%%%%%%%%%%%%%%%%%%%%%%%%%%%
%%     In Ref.~\cite{Spergel:2006hy}, they have used the notation 
%%      {\cal P}_{\rm curv} = | \Del R |^{2} = 2.95 \times 10^{-9} A,
%%      while the curvature perturbation {\cal P}_{\rm curv} is given by 
%%      {\cal P}_{\rm curv} = \frac{1}{12 \pi^{2}} \frac{V^{3}}{V'^{2}}.
%%     The relation between the density fluctuation and the curvature fluctuation
%%      is given by, 
%%      \del^{2} = \frac{4}{25}  {\cal P_{\rm curv}}.
%%%%%%%%%%%%%%%%%%%%%%%%%%%%%%%%%%%%%%%%%%%

\begin{figure}[t!]
\begin{minipage}{0.5\hsize}
 \begin{center}
 \includegraphics[width=6.5cm]{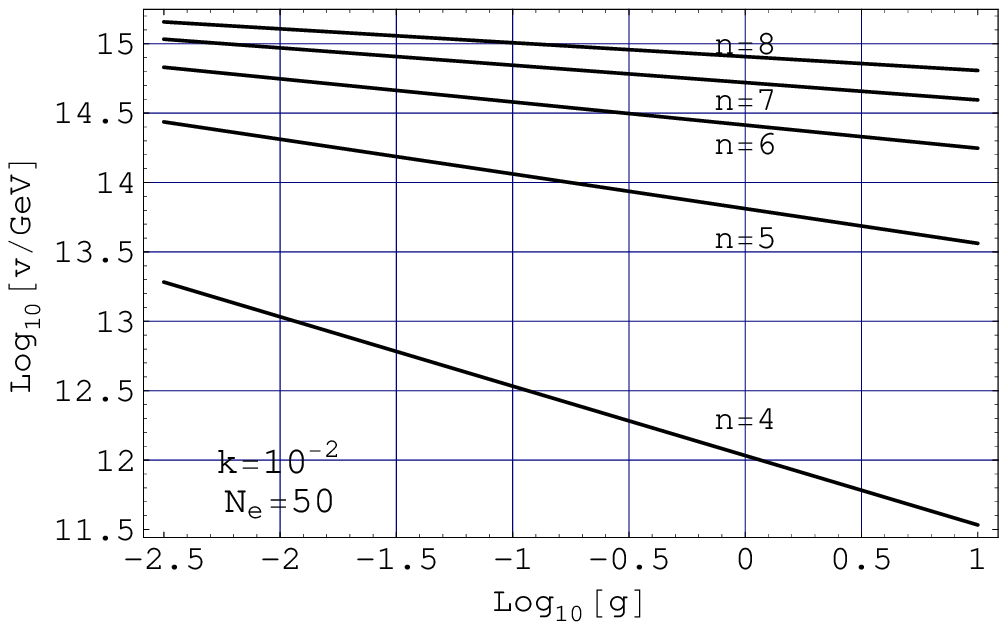}
 \end{center}
 \end{minipage}
 \begin{minipage}{0.5\hsize}
 \begin{center}
 \includegraphics[width=6.5cm]{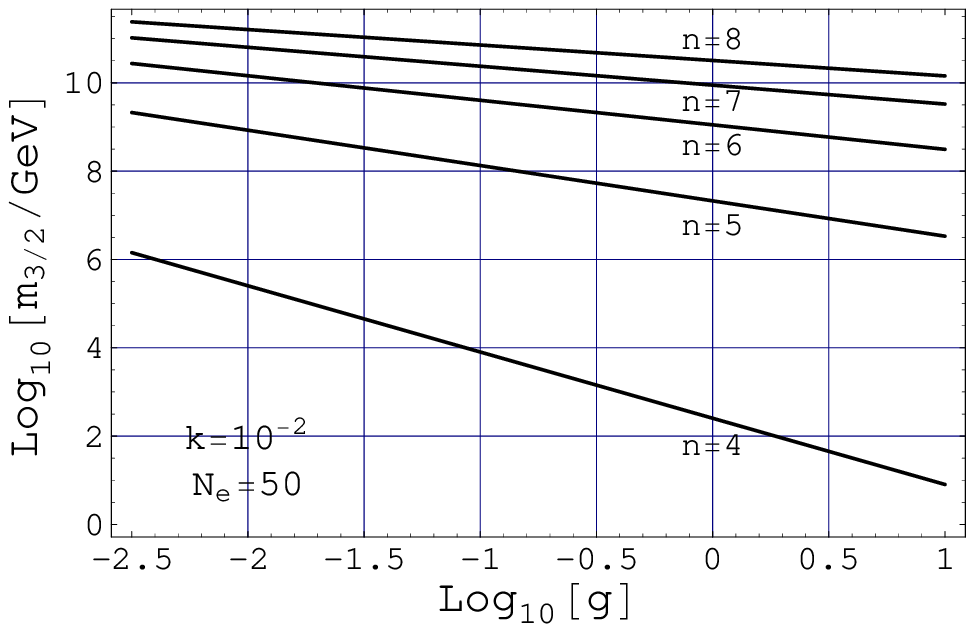}
 \end{center}
 \end{minipage}
  \caption{Left: The $g$ dependence of the inflation scale $v$ for $n=4-8$,  $N_{e}=50$ and 
  $k=10^{-2}$.  
The dependence of $v$ on $N_{e}$ and $k$ are very small.
Right: The $g$ dependence of the gravitino mass for $n=4-8$, $N_e=50$ and $k = 10^{-2}$.
}
 \label{fig:v}
\end{figure}

The left panel of Fig.~\ref{fig:v} shows the $g$ dependence of the inflation scale 
$v$ in Eq.~(\ref{eq:appv}) for $n=4-8$ and a given e-folding number $N_e = 50$.
We see from the figure that the inflation scale increases as the coupling constant $g$
decreases.
We also see that the inflation scale  becomes higher for larger $n$.

The e-folding number $N_{e}$ is related to the inflation scale $v$ and the reheating temperature $T_{R}$ as
\begin{eqnarray}
\label{eq:efold}
N_{e} \simeq 67 + \frac{1}{3} \ln \frac{v^{2}}{\sqrt{3} }  + \frac{1}{3}\ln T_{R}.
\end{eqnarray}
Thus, we can also obtain the inflation scale $v$ for a given $T_{R}$,
instead of for a given $N_{e}$, by re-solving Eq.~(\ref{eq:COBE}), although 
the resultant $T_{R}$ dependence of $v$ is very weak.

%%%%%%%%%%%%%%%%%%%%%%%%%%%%%%%%%%%%%%%%%%%
%%%%%%%%%%%%%%%%%%%%%%%%%%%%

For a given inflation scale $v$ and $n$, we find the gravitino mass from Eq.~(\ref{eq:gmass}).
The right panel of Fig.~\ref{fig:v} shows the $g$ dependence of the gravitino mass
for $n=4-8$. 
Similarly to the inflation scale $v$, the gravitino mass also increases as the $g$ decreases
and as  $n$ increases.
Especially, the gravitino mass  is roughly given by
\begin{equation}
\label{eq:gmassapp}
 m_{3/2}  \sim 300\, g^{-3 / 2} \,{\rm GeV},
 \label{eq:gmass4}
\end{equation}
for $n=4$ , while $m_{3/2}\gsim 10^{6}$\,GeV for $n\geq 5$ and $g<{\cal O}(1)$.
In the following, we fix $n=4$, since we are interested in TeV-scale SUSY breaking.

Interestingly, as we see from Eq.~(\ref{eq:gmassapp}), 
the new inflation model with $n=4$ easily accommodates 
 anomaly-mediation models which are realized for $m_{3/2}\gsim 30$\,TeV,
by taking $g={\cal O}(10^{-2})$.
Furthermore, as we will show in the next section, the inflaton decay may 
provide a sufficient amount of winos which  explains
the observed dark matter density in the universe.

Let us now discuss a reheating process of the present new inflation.
We introduce the following
superpotential interaction between the inflaton and 
the right-handed neutrinos $N$'s,
\begin{equation}
 \delta W = \frac{h}{6}\phi^{3}N^2,
 \label{eq:hNN}
\end{equation}
where $h$ is a dimensionless parameter and we take it positive~\cite{Ibe:2006fs}.
Notice that we have introduced  another parameter $h$ in the model.
At the vacuum, this term Eq.~(\ref{eq:hNN}) induces masses of the right handed neutrinos as
\begin{equation}
 m_N=\frac{h}{3} \phi_0^{3}\simeq \frac{h}{12g} m_{\varphi}.
\end{equation} 
Then, if  $2m_{N}<m_{\varphi}$ (i.e., $h<6g$) the inflaton decays into a pair of right-handed neutrinos 
and the reheating occurs after the inflation.
The decay rate is given by
\begin{equation}
  \Gamma_N \simeq \frac{|h|^2}{16\pi}\phi_0^{4}m_\varphi.
\label{eq:gammaphi}
\end{equation}
Consequently, the reheating temperature
becomes
\begin{equation}
T_R \simeq  \left(\frac{10}{g_*\pi^2} \Gamma_N^2\right)^{\frac{1}{4}} \simeq 
1.5 \times 10^6\, h g^{-5/4}\, \mathrm{GeV},
\label{eq:tr}
\end{equation}
where  $g_*(\simeq 228.75)$ is the effective number of massless degrees of
freedom, and we have used Eqs.~(\ref{eq:phimass}), (\ref{eq:appv}) and (\ref{eq:gammaphi}).

A nice point of this reheating process
is that the production of right-handed neutrinos via the inflaton decay 
causes
the leptogenesis~\cite{Fukugita:1986hr} which results in the baryon asymmetry in the universe.
As investigated in Ref.~\cite{Ibe:2005jf}, the baryon asymmetry per entropy density $s$ is given by,
\begin{equation}
\eta = \frac{n_B}{s} \simeq 8.2 \times 10^{-11} \left(\frac{T_R}{10^6\mathrm{GeV}}\right)\left(\frac{2m_N}{m_\varphi}\right)
 \left(\frac{m_{\nu_3}}{{0.05\mathrm{eV}}}\right)\delta_{eff}.
 \label{lepto}
\end{equation}
Here $m_{\nu_3}$ is the mass of the heaviest (active) neutrino,
the phase $\delta_{eff}={\cal O}(1)$ is the effective CP-violating phase defined in
Ref.~\cite{Fukugita:1986hr}, and  we have assumed the ratio of the vacuum
expectation value of up- and down-type Higgs bosons to be  much larger than 1.
Since $T_{R}$, $m_{\varphi}$ and $m_{N}$ are given by $g$ and $h$, the baryon asymmetry 
can be expressed in terms of  $g$ and $h$, or equivalently $m_{3/2}$ and $T_{R}$, neglecting 
the weak dependence on $k$.

\begin{figure}[t!]
 \begin{center}
 \includegraphics[width=6.5cm]{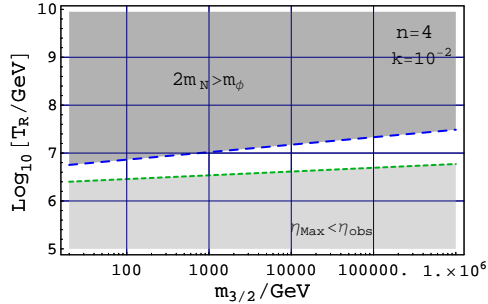}
 \caption{The gravitino mass dependence of the upper  and lower bound on the 
 reheating temperature.
 The dashed (blue) line corresponds to the upper bound on the reheating temperature
 and the short dashed (green) line to the lower bound on the reheating temperature. 
 }
 \label{fig:TR}
 \end{center}
\end{figure}

Fig.~\ref{fig:TR} shows the allowed parameter region on ($m_{3/2}$, $T_{R}$) plane.
The dashed (blue) line is the upper bound on the reheating temperature which
comes from the condition,
\begin{eqnarray}
 m_{N} < \frac{m_{\varphi}}{2}, \qquad (i.e.,\, h<6g).
\end{eqnarray}
The short dashed (green) line denotes the lower bound on the reheating temperature
to explain the observed baryon asymmetry $\eta = (8.7\pm 0.3)\times 10^{-11}$~\cite{Spergel:2006hy}.
As we see from the figure, the allowed region of the reheating temperature is
rather constrained,
\begin{eqnarray}
T_{R} \simeq 10^{6-7}\,\rm GeV.
\end{eqnarray}

%%%%%%%%%%%%%%%%%%%%%%%%%%%%%%%%%%%%%%%%%%%
%%%%%%%%%%%%%%%%%%%%%%%%%%%%%%%%%%%%%%%%%%%

Before closing this section, we summarize the relation between model parameters and the 
physical quantities.
As we see in Eqs.~(\ref{eq:gmass}),  (\ref{eq:spectrum}), (\ref{eq:appv}) and (\ref{eq:tr}), 
all the parameters in the model, $k$, $v$, $g,n$ and $h$, are determined by
the spectral index, the density fluctuations, the gravitino mass and 
the reheating temperature (equivalently the observed baryon asymmetry).
As we have seen so far, four out of the above five parameters are already constrained by 
observations. 
Therefore, there remains no free parameter other than the gravitino mass which is 
one of the most important parameter of the low energy SUSY models and will be determined  by
future experiments.

\section{Dark matter density in anomaly-mediated SUSY breaking}
As mentioned in the introduction we adopt anomaly-mediated SUSY breaking where
gauginos acquire SUSY-breaking masses through anomalies of the scale 
invariance~\cite{Randall:1998uk,Giudice:1998xp}, while
squarks and sleptons receive SUSY breaking soft mas?ses directly from SUGRA effects.
Thus, there are mass gaps between gauginos and squarks /sleptons~\cite{Giudice:1998xp}.

In anomaly-mediation models, the most probable lightest supersymmetry particle (LSP) 
candidate is the neutral wino%
\footnote{
The finite one-loop corrections to the gaugino masses from the Higgs and Higgsino exchanges 
can be non-negligible  when the supersymmetric Higgs mass $\mu$ is of the
same order of $m_{3/2}$~\cite{Giudice:1998xp}.
In the following, we neglect such one-loop corrections.
} which has a mass
 $m_{\tilde w}\simeq 3\times 10^{-3}m_{3/2}$ and it is a good candidate for the dark
 matter~\cite{Giudice:1998xp,Gherghetta:1999sw,Moroi:1999zb,Ibe:2004gh}.%
\footnote{The following discussion is almost independent of  masses of  sfermions as long as 
the LSP is the neutral wino.}
However, the thermally produced winos cannot explain the observed dark matter density unless 
it is very heavy, $m_{\tilde{w}} \simeq 2$\,TeV, because of its large annihilation cross section.
The number density of winos produced via the decay of the gravitinos 
depends on the relic number density of gravitinos.
We easily find that the gravitinos produced by scattering processes in 
the thermal bath cannot supply the observed dark matter density 
unless $T_{R}\gsim 10^{8}$\,GeV.
Thus, if we consider the wino dark matter with $T_{R}\simeq 10^{6-7}$\,GeV, 
the mass of the wino is too heavy ($m_{\tilde{w}}\simeq 2$\,TeV) 
to be found in the next generation of collider experiments.
Fortunately, in our case, there is another source of  the winos,
i.e., the gravitinos produced directly from the inflaton decay.

To show it explicitly, we derive the relic density of gravitinos produced directly 
from the inflaton decay.
The relevant terms of the inflaton decay into a pair of gravitinos are,
\begin{eqnarray}
 K = |S|^{2} + |\phi|^{2} + b |S|^{2}|\phi|^{2} + \cdots,
\label{eq:HiddenInflaton}
\end{eqnarray}
where $S$ is the hidden sector field which has a  non-vanishing $F$-term,
 $b$ a real constant, and the ellipses the higher dimensional terms.
Since we have no symmetries to suppress $b$, we naively expect it to be of order one.

As discussed in Ref.~\cite{Endo:2006tf}
the hidden sector fields and the inflaton $\phi$ mix each other and can be made diagonal by the transformation,
\begin{eqnarray}
\hat{\phi} \simeq \phi + \e  S, \qquad\,\,\,\,\hat{S}  \simeq S - \e^{*} \phi,
\end{eqnarray}
with a mixing angle $\e$,
\begin{eqnarray}
\e \simeq \sqrt{3} (1+b) \phi_{0} \frac{m_{3/2}m_{\varphi}}{m_{S}^{2}}.
\label{eq:mixing}
\end{eqnarray}
Here, we have assumed the vacuum expectation value of $S$ and the holomorphic 
mixing mass terms are negligible.
Note that we have also assumed that mass of the hidden field, $m_{S}$, is much larger than
the inflaton mass $m_{\varphi}\simeq 10^{10}$\,GeV.
This assumption is reasonable for the dynamical SUSY breaking models where
the hidden sector fields have masses of order of the SUSY breaking scale 
$\L_{\rm SUSY}\simeq 10^{12}$\,GeV.

The mixing between hidden and inflaton fields leads to an effective coupling
of the inflaton to the gravitinos~\cite{Endo:2006tf},
\begin{eqnarray}
\label{eq:effcoup}
 |{\cal G}_{\hat\phi}^{(\rm eff)}| \simeq \sqrt{3}\,\, \frac{m_{S}^{2}}{m_{\varphi}^{2}}\,\, |\e|
\simeq  3 \times \frac{m_{3/2}}{m_{\varphi}} (1+b)\phi_{0},
\end{eqnarray}
which induces the decay of $\hat\phi$ into a pair of gravitinos.
The decay rate is given by
\begin{eqnarray}
 \Gamma_{3/2} 
 \simeq \frac{ |{\cal G}_{\hat\phi}^{(\rm eff)}|^{2}}{288\pi} \frac{m_{\varphi}^{5}}{m_{3/2}^{2}M_{G}^{2}}.
\end{eqnarray}
Then,  the gravitino-entropy ratio (yield) is given by~\cite{Kawasaki:2006gs,Kawasaki:2006hm},
\begin{eqnarray}
\label{eq:yield0}
Y_{3/2}^{(\rm inf)} = 2 \frac{\Gamma_{3/2}}{\Gamma_{N}} \frac{3 T_{R}}{4 m_{\varphi}} 
               \simeq 4.5 \times  |{\cal G}_{\hat\phi}^{(\rm eff)}|^{2} 
                \left(\frac{m_{\varphi}}{10^{9}\,\rm GeV} \right)^{4}
                \left(\frac{10^{7}\,\rm GeV}{T_{R}} \right)
                \left(\frac{\rm TeV}{m_{3/2}} \right)^{2}.
\end{eqnarray}
By substituting Eq.~(\ref{eq:effcoup}) into Eq.~(\ref{eq:yield0}), we obtain the yield as,
\begin{eqnarray}
\label{eq:yield}
 Y_{3/2}^{(\rm inf)} &\simeq& 7 \times 10^{-16}   (1+b)^{2}\left(\frac{m_{\varphi}}{10^{9}\,\rm GeV} \right)^{2}
                \left(\frac{10^{7}\,\rm GeV}{T_{R}} \right)
                \left(\frac{\phi_{0}} {10^{16}\rm GeV} \right)^{2},\\
                &\simeq&  
               4 \times 10^{-13} \times (1+b)^{2}\left(\frac{m_{3/2}}{100\,{\rm TeV}}\right)^{4/3} \left(\frac{10^{7}\,\rm GeV}{T_{R}} \right),
 \end{eqnarray}
where we have also used Eqs.~(\ref{eq:phivev}), (\ref{eq:phimass}), (\ref{eq:appv}) 
and (\ref{eq:gmass4}).
We see that there are no mass suppression factors discussed in Ref.~\cite{Endo:2006tf,Dine:2006ii},
since the mass of the hidden sector field $S$ is larger
than that of the inflaton.
Notice that we have a substantial amount of gravitinos even for the minimal K\"ahler potential, i.e.,
for $b=0$.

The above gravitinos decay to the winos, and the resultant yield of the winos is given by%
\footnote{The annihilation process of the non-thermally produced winos is ineffective,
since the gravitino decay occurs at very low temperatures, $T \simeq {\cal O}(10)$\,MeV. }
\begin{eqnarray}
\label{eq:winoinf}
 Y_{\tilde{w}} \simeq Y_{3/2}^{(\rm inf)}.
\end{eqnarray}
There are other contributions to the yield of the winos, one from the thermally
produced winos,
\begin{eqnarray}
\label{eq:winoth}
Y_{\tilde w} \simeq 10^{-14}\left(\frac{m_{\tilde w}}{100\,{\rm GeV}}\right),
\end{eqnarray}
and the other from the decay of gravitinos produced by the thermal 
scattering~\cite{Kawasaki:1994af,Bolz:2000fu},
 \begin{eqnarray}
 \label{eq:winoscatt}
 Y_{\tilde w} \simeq 2\times 10^{-15} \left(\frac{T_{R}} {10^{7}\,\rm GeV} \right).
\end{eqnarray}
By comparing Eqs.~(\ref{eq:winoinf})-(\ref{eq:winoscatt}), we find that 
the dominant source of the winos is gravitinos produced directly from 
the inflaton decay for $T_{R}\simeq 10^{6-7}$\,GeV.
Then, we find that the mass density parameter of the wino is given by
\begin{eqnarray}
  \Omega_{\tilde w}h^{2} &=& \frac{m_{\tilde w} Y_{\tilde w}}{3.5 \times 10^{-9}\,{\rm GeV}},\\
  &\simeq&  0.04 \times (1+b)^{2} \left(\frac{m_{3/2}}{100\,{\rm TeV}}\right)^{7/3}
  \left(\frac{10^{7}\,\rm GeV}{T_{R}} \right),
\end{eqnarray}
for $T_{R}\simeq 10^{6-7}$\,GeV.

\begin{figure}[t]
\begin{minipage}{0.5\hsize}
 \begin{center}
 \includegraphics[width=8cm]{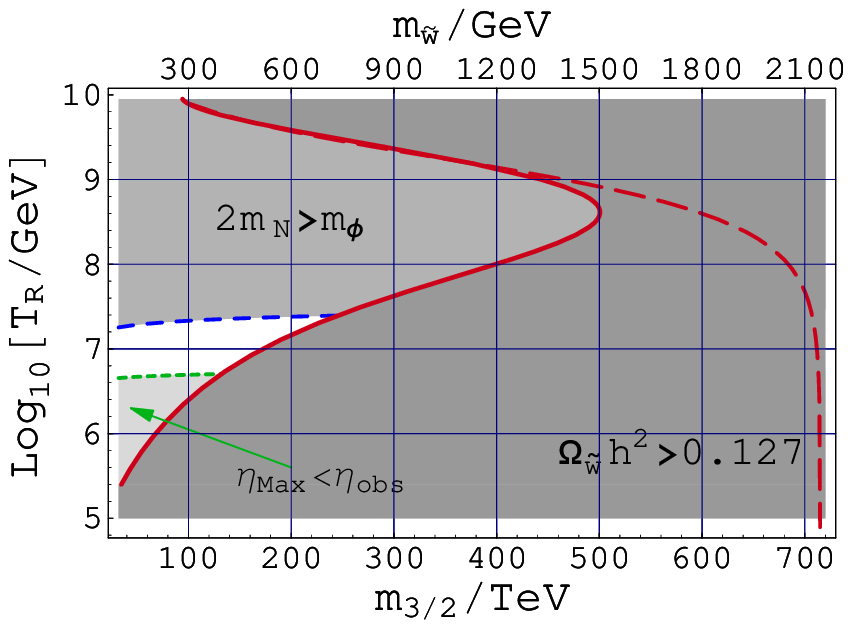}
 \end{center}
 \end{minipage}
 \begin{minipage}{0.5\hsize}
 \begin{center}
 \vspace{.5cm}
 \includegraphics[width=7cm]{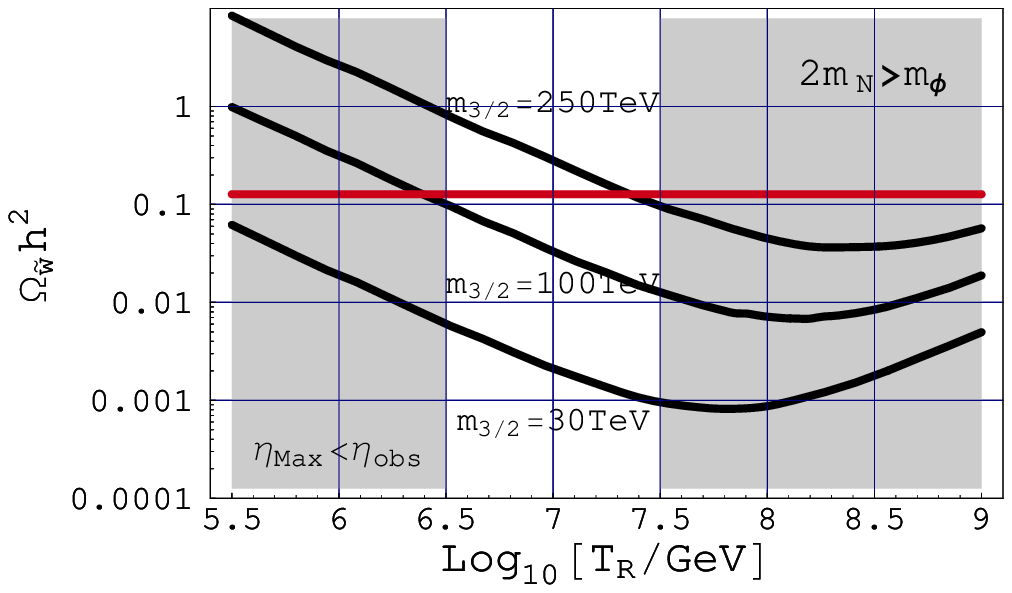}
 \end{center}
 \end{minipage}
  \caption{Left: The $\Omega_{\tilde{w}}h^{2}\simeq 0.127$ region on the ($m_{3/2},T_{R}$) plane.
 The solid (red) line corresponds to the wino abundance $\Omega_{\tilde w}h^{2}\simeq 0.127$ 
 and the long-dashed line to the wino abundance without the inflaton contribution discussed
 in the text.
Right: The $T_{R}$ dependence of the wino mass density  $\Omega_{\tilde w}h^{2}$.
The solid lines correspond to  $m_{3/2} =30\,{\rm TeV}$, $100\,{\rm TeV}$ and $250\,{\rm TeV}$
from the bottom to up, and the thin (red) line to $\Omega_{\tilde w}h^{2}\simeq 0.127$.
In both panels, we have used $b=0$, for simplicity.
}
 \label{fig:DM}
\end{figure}

Fig.~\ref{fig:DM} shows the mass density of the wino for $b=0$.
We plot the wino mass density in a ($m_{3/2}$, $T_{R}$)  plane in the left panel, 
and the $T_{R}$ dependence of the wino density for a given gravitino mass in the right panel. 
In the figures, we have also scanned  $k$ from  $10^{-1.5}$ to $10^{-4}$,
although the resultant $k$ dependence is negligibly small.

The solid (red) line in the left panel corresponds to the observed dark matter density
$\Omega_{\rm DM}h^{2} \simeq 0.127$, the right side of the line to the region of too 
much dark matter $\Omega_{\tilde w}h^{2}>0.127$, 
and the left side of the line to the region of less dark matter $\Omega_{\tilde w}h^{2}<0.127$.
For comparison, we also plot the line where $\Omega_{\tilde w}h^{2} \simeq 0.127$ is satisfied
without the gravitinos directly from the inflaton decay as a long-dashed (red) line in the left panel.
This line corresponds to the result in Ref.~\cite{Ibe:2004gh}, although 
they have used $m_{\tilde w} \simeq 5.2\times 10^{-3} m_{3/2}$ because of the
existence of the light charged particles.
From the figures, we find that the mass density of the wino is consistent with the observed dark matter
density in a narrow range of the wino mass, $400$\,GeV$\lsim m_{\tilde w}\lsim 750$\,GeV, 
for $T_{R}\simeq 10^{6-7}$\,GeV and $b=0$.

As a result, we find that the gravitinos directly from the inflaton decay provide a sufficient amount
of the winos for the dark matter if
\begin{equation}
(1+b)^{-6/7}\times 400\,{\rm GeV}\lsim m_{\tilde w}\lsim 
{\rm Min}\left[(1+b)^{-6/7}\times 750\,{\rm GeV},
\,\, 2.1\,{\rm TeV}
\right].
\end{equation}
Here, the upper  bound on the wino mass on the right hand side, $m_{\tilde w}\simeq 2.1$\,TeV,
corresponds to  the long-dashed (red) line in Fig.~\ref{fig:DM} where the observed dark matter density 
is supplied by the thermally produced winos.
Since the parameter $b$ is of  order one, we consider that the $|b|$ ranges from 
$1/3$ to $3$.
In that case, the mass density of the wino is consistent with the observed dark matter
 in a wide range of the wino mass, $100$\,GeV$\lsim m_{\tilde{w}} \lsim2$\,TeV.

%%%%%%%%%%%%%%%%%%%%%%%%%%%%%%%%%%%%%%%%%%%
%%%%%%%%%%%%%%%%%%%%%%%%%%%%%%%%%%%%%%%%%%%

\section{Conclusions}
In this paper, we study the new inflation model~\cite{Kumekawa:1994gx,Izawa:1996dv}
 which is well consistent with the WMAP observations.
The remarkable feature of this  model is that there remains essentially 
only one free parameter,
the gravitino mass, and the other parameters are determined by the observations.
However, as discussed in Refs.~\cite{ISY2}, gravity-mediation models for SUSY breaking 
suffer from a serious gravitino-overproduction problem, and hence, the new inflation model
suggests gauge-mediation or anomaly-mediation models.

As to gauge-mediation models, however, they do not well accord with the new inflation
model compared to the anomaly-mediation models.
Firstly, the gravitino mass in this new inflation model is rather large, i.e., 
$m_{3/2} \geq{\cal O}(10)$\,GeV,
where the equality is saturated for a large coupling constant $g\simeq 10$ 
(see Eq.~(\ref{eq:gmass4})).%
\footnote{The coupling constant $g$ should be at most $g\lsim 10$, since otherwise the large 
coupling constant leads to large radiative corrections to the K\"ahler potential which 
invalidates the flatness of the inflaton potential.}
Secondly, the gravitino is the LSP, but it is difficult to explain the observed dark matter density,
since the gravitino abundance is in short supply for $T_{R}\simeq 10^{6-7}$\,GeV and 
$m_{3/2}\simeq{\cal O}(10)$\,GeV (see Eq.~(\ref{eq:yield}).).

On the contrary, anomaly-mediation models are in harmony with the new inflation model.
As we have shown, the new inflation model with $n=4$ can easily realize the gravitino mass
$m_{3/2}\gsim 30$\,TeV which is suitable for the anomaly-mediation.
Furthermore, we have found that gravitinos produced directly from the inflaton
decay provide a sufficient amount of the winos for the dark matter in the universe.
As a result, we found that the relic density of the wino LSP is consistent 
with the observed dark matter density in a wide range
of the wino mass, $100$\,GeV$\lsim m_{\tilde{w}} \lsim2$\,TeV.

\section*{Acknowledgments}
M.~I. thanks the Japan Society for the Promotion of Science
for financial support. This work is partially supported by Grand-in-Aid Scientific Research (s) 14102004.
The work of T.T.Y. has been supported in part by a Humboldt Research Award. 

\end{document}